\documentclass[useAMS,usenatbib]{mnras}

\usepackage{natbib}
\usepackage{graphicx, setspace, subfig, latexsym, amssymb, amsmath, booktabs, amsfonts, wasysym, paralist}
\usepackage{multirow}
\usepackage{gensymb}
\usepackage{mathrsfs}

\usepackage[T1]{fontenc}
\usepackage{ae,aecompl}
\usepackage{txfonts}

\title[Searching for local host galaxies of RRATs]{A novel approach for identifying host galaxies of nearby FRBs}

\author[A. Rane and A. Loeb]{A. Rane$^{1}$\thanks{email: arane@mix.wvu.edu} and A. Loeb$^{2}$\thanks{email: aloeb@cfa.harvard.edu} \\
$^{1}$Department of Physics \& Astronomy, West Virginia University, Morgantown, WV, 26506 USA\\
$^{2}$Astronomy Department, Harvard University, Cambridge, MA 02138, USA\\
}

\date{\today}

\pubyear{2016}

\begin{document}
\label{firstpage}
\pagerange{\pageref{firstpage}--\pageref{lastpage}}
\maketitle

\begin{abstract}
We report on a search for host galaxies of a subset of Rotating Radio Transients (RRATs) that possess a dispersion measure (DM) near or above the maximum Galactic value in their direction. These RRATs could have an extragalactic origin and therefore be Fast Radio Bursts (FRBs). The sizes of related galaxies on the sky at such short distances are comparable to the beam size of a single-dish telescope (for example, the $7.0'$ radius of the Parkes beam). Hence the association, if found, could be more definitive as compared to finding host galaxies for more distant FRBs. We did not find any host galaxy associated with six RRATs near the maximum Galactic DM. This result is consistent with the fact that the probability of finding an FRB host galaxy within this volume is also very small. We propose that future follow-up observations of such RRATs be carried out in searching for local host galaxies as well as the sources of FRBs. 
\end{abstract}




\section{Introduction}
\label{sec:intro}
Fast Radio Bursts (FRBs) are millisecond-duration radio signals, first reported in 2007 \citep{lorimerburst2007} and since then detected by the Parkes, Arecibo, and Green Bank telescopes between $0.7-1.5~ \mathrm{GHz}$, primarily through processing of pulsar surveys and a few in the real-time detection pipeline. To date, 17 FRBs have been published\footnote{See http://www.astronomy.swin.edu.au/pulsar/frbcat/} (\citealt{lorimerburst2007}; \citealt{keanefrb2012}; \citealt{thorntonfrb2013}; \citealt{spitlerfrb2014}; \citealt{burkespolaorfrb2014}; \citealt{petrofffrb}; \citealt{ravifrb2015}; \citealt{championfrb2016}; \citealt{masui2015}; \citealt{keanehost2016}) and two of them had been observed to repeat (\citealt{scholzrepeat2016}; \citealt{maoz2015}). Fourteen of these sources have been detected at high Galactic latitudes ($|b|>5\degree$) and their large dispersion measures (DM $\sim 375-1700~ \mathrm{pc~ cm^{-3}}$) exceed the expected Galactic contribution predicted by the NE2001 model in the direction of the bursts \citep{cordesne2001}, suggesting an extragalactic origin. However, no host galaxy has yet been confirmed for any of these events. \citet{keanehost2016} proposed association of FRB~150418 with an elliptical galaxy at redshift 0.5; however, \citet{wb2016} and \citet{vedantham2016} have demonstrated that the radio afterglow emission is instead due to AGN variability. Since the telescopes that are efficient in detecting these bursts are single-dish, the beam sizes are typically a few arcminutes which makes the position uncertainty large, limiting the ability to identify the host galaxy. The excess DM suggests redshifts in the range $\sim 0.3-1.3$ for FRBs if it originates from the intergalactic medium. Given these circumstances, it is extremely difficult to associate an FRB with a particular host galaxy unless there is some extraordinary evidence for coincident, transient multiwavelength emission within the beam. 
Another class of transients, known as RRATs are a group of Galactic pulsars that emit sporadic pulses \citep{maurarrats2006}. There are about 112 RRATs discovered so far and some of them have been observed at only one epoch or have been observed to emit only one pulse\footnote{See RRATalog: \\http://astro.phys.wvu.edu/rratalog/}. Before it was discovered that FRB~121102 is repeating, \citet{keane2016} discussed the uncertainty in the RRAT/FRB classification by analysing RRATs from which only one pulse has been detected so far. But since we know of repeating FRBs, there is no distinction between pulses from RRATs and FRBs. The only clear distinction is associated with their DM values. So, finding a host galaxy is really important at this point in order to constrain emission models for these bursts. 

Since pulses from RRATs are essentially indistinguishable from FRBs, as explained in detail in \S~2, we decided to find and confirm the origin of a subset of RRATs that are close to the edge of the galaxy. Taking into account the uncertainty in the free electron density model, these RRATs could possibly have an extragalactic origin. The excess DM for our sample is not as high as seen in FRBs, corresponding to very low redshifts at which the size of the host galaxy on the sky is comparable to the beam size of a single-dish telescope (for example, the Parkes beam has a HPBW of 14'). This similarity of scales would make the association, if found, more definitive than at cosmological distances. 

In \S~2 we compare the most commonly used model for the Galactic distribution with the newly available model to derive the DM distribution of radio transients on the sky. In \S~3 we outline the basic criterion used to choose our RRAT sample. The results for each RRAT candidate are discussed in \S~4 . In \S~5 we discuss the non-detection of host galaxies and related probabilities. Finally, in \S~6 we summarize our results and present our conclusions.
 
\section{Galactic free electron density models}\label{sec:models}
The NE2001 model \citep{cordesne2001} describes the structure of ionised gas in the Galaxy and is widely used to estimate distances to radio pulsars for which DM is the only distance indicator. In the case of FRBs for which the DMs are too high, this model is used to estimate the DM contribution in the direction of FRB from the Galaxy. This model is based on the observed DMs of Galactic radio pulsars and includes contributions from the thin disk associated with low-latitude HII regions, the thick-disk, the spiral arms, small-scale features corresponding to local ISM, individual high-density clumps and voids. However, the uncertainty in the NE2001 model is about $\sim 20\%$, particularly at higher latitudes, as also discussed in \citet{gaensler2008}. 

\begin{figure}
\centering
\includegraphics[scale=0.7]{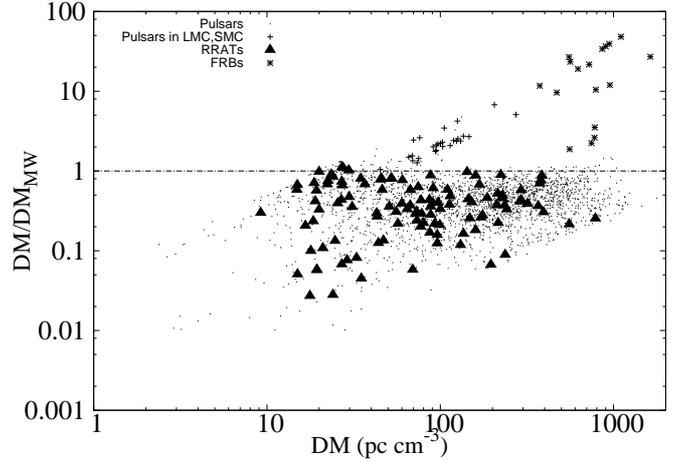}
     \caption{The ratio of measured DM to maximum Galactic DM versus the measured DM (in $\mathrm{pc~ cm^{-3}}$) for all radio transients. The maximum Galactic DM is calculated by integrating the YMW16 model to the edge of the Galaxy for each transient's direction. The dashed line shows the maximum ratio expected for Galactic objects if the electron density is accurate for all lines of sight.}
     \label{fig:dm-ymw16}
\end{figure}

Recently, a new model by Yao, Manchester and Wang\footnote{http://www.atnf.csiro.au/research/pulsar/ymw16/}, called as YMW16 has been proposed for the distribution of free electrons in the Galaxy, the Magellanic Clouds, and the inter-galactic medium (IGM). This model is based on measurements from 189 pulsars with independently determined distances as well DMs. We compared the two models by integrating both models to the edge of the Galaxy for each radio transient's direction. The list of pulsars in the Milky Way, SMC and the LMC is obtained from \citet{atnf2005}. The  ratio ($r$) of the measured DM to the maximum Galactic DM versus the measured DM is plotted for the NE2001 model in \citet{spitlerfrb2014}. We show a similar plot for the YMW16 model in Figure \ref{fig:dm-ymw16}. We can see that the galactic DM contribution is lowered along certain lines of sight towards the galactic center (GC) pulsars, minimizing the gap between GC pulsars and the overall pulsar population. Pulsars in the LMC and SMC and FRBs, all have $r>1$, consistent with the NE2001 model. A small fraction of pulsars have $r>1$ in both models, possibly due to uncertainties or them being in the Galactic halo but we will not discuss these pulsars here. Another promising difference between the two models is that all RRATs had $r<1$ according to the NE2001 model, thus confirming their Galactic origin; however, according to the YMW16 model, some RRATs now have shifted to having $r>1$. Assuming this model is closer to the true values than the NE2001 model, these RRATs are very similar to FRBs. This RRAT sample provides a promising opportunity to find host galaxies that are close to us since their excess DM is much lower than most of the FRBs.

\begin{figure*}
\centering
\includegraphics[scale=0.4,angle=270]{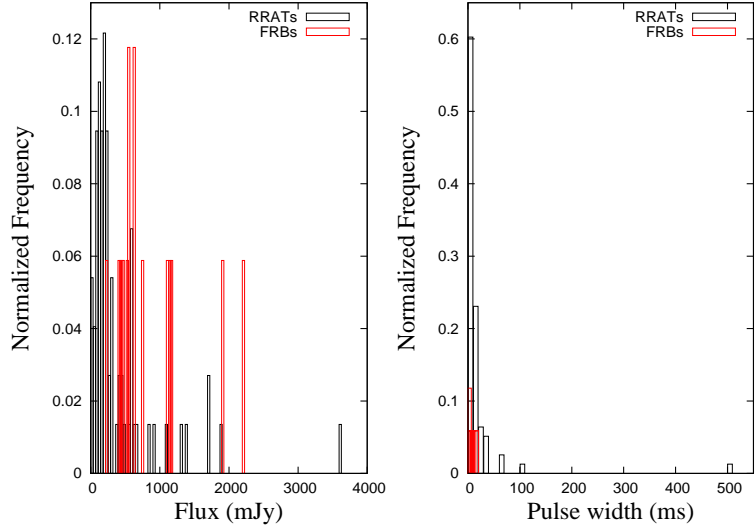}
     \caption{\textbf{The normalized histograms of peak fluxes and pulse widths of RRATs and FRBs.}}
     \label{fig:hist}
\end{figure*}

We also compared the pulse width and flux distributions for both RRATs and FRBs, as seen in Figure \ref{fig:hist}. The two sample Kolmogorov-Smirnov test gives $\mathrm{D=0.37}$ and a p-value of 0.03 for the pulse widths, thus indicating that the pulse width distributions of the two populations are not significantly different. For the flux distribution, $\mathrm{D=0.80}$ and the p-value is small ($9.1\times10^{-9}$). This is not surprising since the peak fluxes of known FRBs are higher than RRATs.

\section{Our sample}
\begin{figure}
\centering
\includegraphics[scale=0.7]{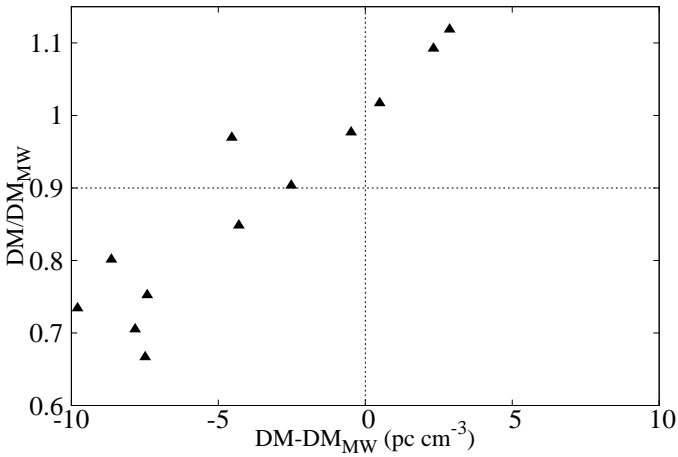}
     \caption{A zoomed in plot of the ratio between the measured DM and the maximum Galactic DM versus measured DM subtracted from maximum Galactic DM for RRATs.}
     \label{fig:dmzoom}
\end{figure}

In an attempt to find host galaxies as discussed in \S~1, we have selected a sample of RRATs which have $r$ greater than 0.9 and for which $\rm{DM_{diff}= DM-DM_{MW}}$ is less than 10 $\rm{pc~ cm^{-3}}$ as seen in Figure \ref{fig:dmzoom}. These RRATs are at the edge of our galaxy and if the uncertainties in the electron density model are taken into account, they could be Galactic or extragalactic.

\section{Results}
\begin{figure*}
\centering
\includegraphics[scale=0.5]{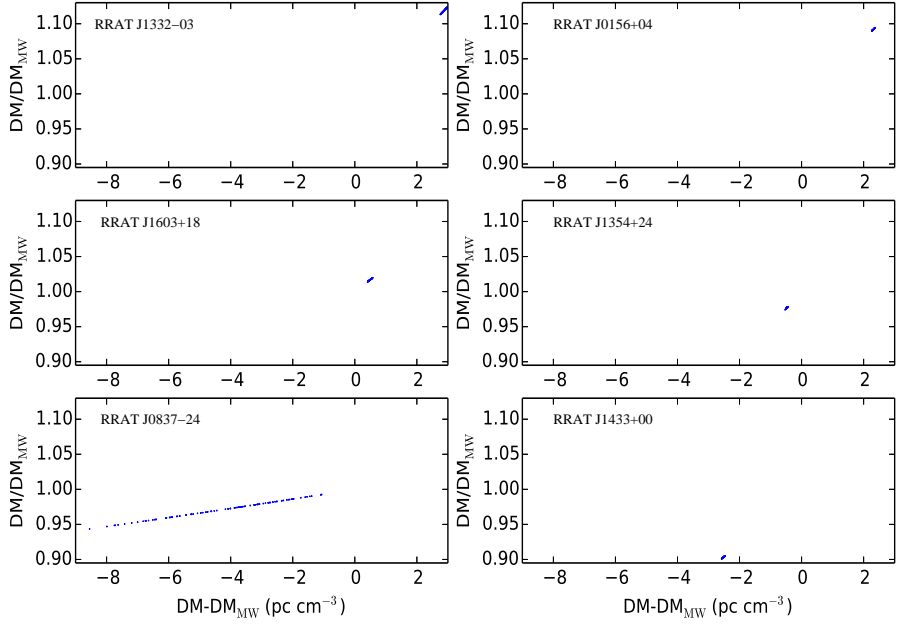}
     \caption{A ratio of measured DM to maximum Galactic DM plotted against the difference in measured DM and maximum Galactic DM for each RRAT in our sample within the beam uncertainty.}
     \label{fig:dmall}
\end{figure*}

Next, we discuss the individual RRAT candidates which could possibly be FRBs. If the DM due to the host galaxy is neglected and if the excess DM$_{\rm diff}$ is assumed to be entirely due to the intergalactic medium, then we infer a redshift $z \sim 0.005$ and distances up to $\sim 20~\rm{Mpc}$. But since these RRATs have low DM and the Galactic $\rm{DM_{MW}}$ uncertainties might be within this excess, we search for galaxies within 120 Mpc (corresponding to a DM $\sim 30 \mathrm{pc~ cm^{-3}}$ for the local density of the intergalactic medium) at the corresponding beam size based on the redshift information provided on NED\footnote{The NASA/IPAC Extragalactic Database (NED) is operated by the Jet Propulsion Laboratory, California Institute of Technology, under contract with the National Aeronautics and Space Administration.}. The summary is given in Table~\ref{table:summaryt}. 
The $\rm{DM_{halo}}$ contribution is determined from the free electron density profile as a function of galecto-centric radius obtained by the latest model that fits best O VIII observations (see, Figure 8 of \citealt{mb2015}). The number of objects found in NED within this search radius are listed in each subsection, and if the spectrum is available then their redshifts are determined by cross-correlating the spectrum against template spectra using the IRAF task \textsc{xvsao} in the \textsc{rvsao} package. We also determined the variation in $\rm{DM_{MW}}$ within the beam uncertainty using the YMW16 model for each RRAT, as seen in Figure~\ref{fig:dmall}. The $\rm{DM_{MW}}$ variation is within $12\%$ for all six RRATs. 
\begin{table*}
  \centering
	\caption{The subset of RRATs included in our sample. Columns 1 to 5 list the RRAT name, Galactic longitude and latitude, uncertainty in position ($\delta$), and measured DM, as obtained from the RRATalog. Columns 6 lists the Galactic contribution to the DM, column 7 lists the ratio of measured DM to the maximum Galactic DM along that line of sight. Column 8 lists contribution to the DM from the Galactic halo and column 9 reports number of extragalactic objects seen within the search radius of beam uncertainty. The search was carried out in NED out to 120 Mpc.} 
        
        \begin{tabular}{ccccccccc}
		\hline
		Name & \multicolumn{1}{c}{$l$} & \multicolumn{1}{c}{$b$} & \multicolumn{1}{c}{$\delta$} & \multicolumn{1}{c}{$DM$} & \multicolumn{1}{c}{$DM_{\rm mw}$} & \multicolumn{1}{c}{$r$} & \multicolumn{1}{c}{$DM_{\rm halo}$} & \multicolumn{1}{c}{$N_{\rm obj}$} \\
	RRAT & \multicolumn{1}{c}{($\degree$)} & \multicolumn{1}{c}{($\degree$)} & \multicolumn{1}{c}{($\mathrm{'}$)} & \multicolumn{1}{c}{($\mathrm{pc~cm^{-3}}$)} & \multicolumn{1}{c}{($\mathrm{pc~cm^{-3}}$)} &  & \multicolumn{1}{c}{($\mathrm{pc~cm^{-3}}$)} & \\ 
		\hline
		J1332$-$03 & 322.25 & 57.91 & 19.4 & 27.1 & 24.23 & 1.12 & 0.5 & 4 \\
		J0156$+$04 & 152.00 & $-$55.00 & 7.5 & 27.5 & 25.18 & 1.09 & 0.5 & 1 \\
	    J1603$+$18 & 32.85 & 45.28 & 7.5 & 29.7 & 29.21 & 1.02 & 0.5 & 0 \\
	    J1354$+$24 & 27.43 & 75.78 & 19.4 & 20.0 & 20.48 & 0.98 & 0.07 & 0 \\
	    J0837$-$24 & 247.45 & 9.80 & 7.0 & 142.8 & 147.3 & 0.97 & 0.22 & 1 \\
	    J1433$+$00 & 349.75 & 53.79 & 7.5 & 23.5 & 26.02 & 0.90 & 0.06 & 1 \\
		\hline
		\end{tabular}
\label{table:summaryt}
\end{table*}
\subsection{RRAT J1332$-$03}
This RRAT was discovered in the 350-MHz Drift-scan pulsar survey with the GBT and was confirmed in follow-up observation. The uncertainty in the beam position is 19.4' at 350 MHz. The four extra-galactic source galaxies found within this search radius are listed in Table ~\ref{table:summaryt2}. For LCRS B133012.2-031854, we did not find any spectrum from online literature. All of these galaxies have higher redshifts than what we would require to account for the intergalactic medium contribution, $\rm{DM_{IGM}}$. Hence these galaxies are most probably not related to this RRAT. 
\begin{table*}
  \centering
	\caption{The extragalactic objects seen within the search radius of beam using NED. For each RRAT, column 2,3, and 4 give the names of the extragalactic sources, their separation from the RRAT in arcminutes ($'$) and redshift ($z$). Column 5 lists corrected redshifts determined from the spectra.} 
        \begin{tabular}{lccccc}
		\hline
		RRAT & Name & \multicolumn{1}{c}{Separation} & \multicolumn{1}{c}{$z$ from NED} & \multicolumn{1}{c}{$z$ from spectra} \\
	     & & \multicolumn{1}{c}{$(')$} & & \\ 
		\hline
	    J1332$-$03 & LCRS B133012.2-031854 & 14.522 & 0.022482 & - \\
	    & GALEXASC J133140.75-030956.2 & 16.692 & $-$0.000076 & 0.063  \\
		& 2dFGRS N138Z073 & 17.365 & $-$0.000200860 & 0.063  \\
	    & 2dFGRS N138Z028 & 18.214 & 0.000200 & 0.055  \\
	    \hline
	    J0156$+$04 & IC 1750 & 5.44 & 0.018860 & 0.018838 \\
	    \hline
	    J1603$+$18 & - & - & - & - \\
	     \hline
	    J1354$+$24 & - & - & - & - \\
	     \hline
	    J0837$-$24 & 2MASX J08374183-2451356 & 0.313 & - & 0.15 \\
	    \hline
	   J1433$+$00 & 2dFGRS N346Z227 & 3.052 & $-$0.000500 & 0.07 \\
		\end{tabular}
\label{table:summaryt2}
\end{table*}
\subsection{RRAT J0156$+$04}
This RRAT was discovered in the single-pulse search of the data obtained in the Arecibo Drift Pulsar survey at 327 MHz and two pulses were observed from it at only one epoch \citep{deneva16}. Follow-up observations detected no pulses from this RRAT. The uncertainties in both the coordinates are $7.5'$, the 327 MHz beam radius. We found one galaxy within this beam with a too high redshift of 0.18, hence indicating no association with this RRAT (Table~\ref{table:summaryt2}).
\subsection{RRAT J1603$+$18}
This RRAT was also discovered in the Arecibo Drift Pulsar survey at 327 MHz and was confirmed in follow-up observation \citep{deneva16}. We did not find any galaxy within a search radius of $7.5'$ from the beam center up to 120 Mpc.  
\subsection{RRAT J1354$+$24}
This RRAT was discovered in the Green Bank North Celestial Cap survey (GBNCC) at 350 MHz\footnote{http://www.hep.physics.mcgill.ca/~karakoc/GBNCC.html}. We did not find any galaxy within a search radius of $19.4'$ up to 120 Mpc.   
\subsection{RRAT J0837$-$24}
This RRAT was discovered in the single-pulse search of the High Time Resolution Universe (HTRU) pulsar survey carried out with the Parkes telescope at $\sim 1.4$~GHZ \citep{burkeIII2011}. We could not find a spectrum for this galaxy and hence the redshift is determined using the luminosity-size relation (see, equation 4 of \citealt{mcintosh2005}). The inferred redshift of 0.15 is too high, indicating no association with this RRAT (Table~\ref{table:summaryt2}). 
\subsection{RRAT J1433$+$00}
This RRAT was discovered in the Arecibo Drift Pulsar survey at 327 MHz and was confirmed in follow-up observation \citep{deneva16}. The 2dFGRS source found within the search radius of $7.5'$ was at an inferred redshift of 0.07, too high than expected, so it is unrelated to the RRAT (Table~\ref{table:summaryt2}).  

\section{Discussion}
Since we did not find any possible host galaxy in \S~4 that might be associated with any of the six RRATs in our sample, these most likely do not have an extragalactic origin. For RRAT J1603$+$18, the DM$_{\rm halo}$ contribution adds to match the measured DM and hence this RRAT might be within the halo of our galaxy. We have also determined the DM associated with the local group by checking the direction of each of these RRATs based on the right panel of Figure 3 in \citet{rl2014}, which yields a DM $\sim 5 \rm{pc~ cm^{-3}}$. RRATs J1332$-$03 and J0156$+$04 could reside within the local group.\newline
The probability of finding a galaxy within the volume out to 120 Mpc by chance for a beam radius of $7.5'$ is $\sim 0.043$, whereas for a beam radius of $19.4'$ it is $\sim 0.28$, based on the average number density of galaxies within the search volume, using NED. Our null result is thus consistent with this estimate. 

\section{Conclusions}
We have presented a search for host galaxies in a subset of RRATs that are at the edge of our Galaxy. These RRATs are interesting since they could either have Galactic or extragalactic origin. In the latter case, the sizes of the host galaxies on the sky at such distances would be comparable to the beam size of a single-dish telescope. We did not find any nearby host galaxy associated with the six RRATs in our sample. Although finding nearby galaxies for such RRATs that could possibly be FRBs is a novel approach, the probability of actually finding a nearby host galaxy is low. Nevertheless, we suggest applying this search strategy to new discoveries of RRATs that will be in the uncertainty zone of the electron density model. Follow-up observations could determine if these RRATs are of a Galactic origin or are extragalactic FRBs, and help us pin down the mysterious origin of FRBs. 

\section*{Acknowledgements}
We thank Dr. Duncan Lorimer for insightful comments and careful reading of the paper. A. Rane would like to thank the hospitality of the Institute for Theory and Computation (ITC). This work was supported in part by NSF grant AST-1312034.

\bsp	
\label{lastpage}
\end{document}